\documentstyle[12pt,aps,epsf]{revtex}
\draft
\preprint{Pramana- J. Phys., Vol 50, No.1, January 1998, pp. 1-9}
\title
{Electrical properties of a-antimony selenide}
\author
{SANJEEV GAUTAM\thanks{e-mail: gautam\%phys@puniv.chd.nic.in}, D K
SHUKLA, SHELLY JAIN and N GOYAL}
\address
{Centre of Advanced Study in Physics,\\
Panjab University, Chandigarh - 160 014 (India).}
\date{\today}
\def\be{\begin{equation}}
\def\beqn{\begin{eqnarray}}
\def\ee{\end{equation}}
\def\eeqn{\end{eqnarray}}
\def\sbse{Sb$_2$Se$_3$}
\def\num{\nonumber}
\def\ba{\begin{array}}
\def\ea{\end{array}}
\begin{document}
\maketitle
\begin{abstract}
This paper reports conduction mechanism in a-\sbse over a wide range
of temperature (238K to 338K) and frequency (5Hz to 100kHz). The d.c.
conductivity measured as a function of temperature shows semiconducting
behaviour with activation energy $\Delta$E= 0.42 eV. Thermally induced
changes in the electrical and dielectric properties of a-\sbse have
been examined. The a.c.  conductivity in the material has been explained
using modified CBH model. The band conduction and single polaron hopping
is dominant above room temperature. However, in the lower temperature
range the bipolaron hopping dominates.
\end{abstract}
\noindent
{\bf PAC Nos ~~~~ 72-20; 71-55 }
\narrowtext
\section{Introduction}
During the last few years amorphous semiconductors have been used in the
manufacture of solar cells, batteries and phototransistors as well as in
some steps of technological processing of very large scale integration
(VLSI) microelectronics circuits \cite{acad86}. Thermally induced
structural and electrical effects are known to be important in inducing
the memory phenomenon in semiconducting chalcogenide glasses
\cite{adp70}-\cite{h70}.  Alzewel et al., \cite{alz75,alz81} have studied the
electrical conductivity of powdered antimony chalcogenides. Antimony
alloys have recieved great attention in the past due to semiconductivity
of Sb concentration. It was reported by Mott and Devis \cite{mott71} that the
effect of Sb in Se is even greater than in Te in promoting 
crystallisation. Recently, study has been made on thin and thick films of
the system $Bi_{10}Sb_xSe_{90-x}$ ($x =$ 35, 40 and 45) \cite{mfadel96}.
Crystalline \sbse has been prepared and studied by several workers
\cite{brch80,fk78}. Antimony triselenide (band gap $\sim$ 1.2 eV),
\cite{wood72} a p-type semiconductor having an orthorhombic $D^5_8$
structure with unit cell dimensions $a = 11.77 \pm 1, b = 3.962 \pm 7, c =
11.62 \pm 1 A^0$ \cite{wa90} is used in preparation of recording
optical laser disc and as an insulating material in MIS solar cells.
The preparation and study of amorphous-\sbse can throw light on the
possible applications of this material in that form. Measurements of
electrical properties of a-\sbse are reported in this paper over a
wide temperature and frequency range. Results indicate that electrical
properties can be explained on the basis of widely accepted
modified-Correlated Barrier Hopping (m-CBH) model for chalcogenide glasses.

\section{Experimental Details}
\label{exp}
Ingots of glassy material \sbse were prepared by melt quench method.
The 99.99\% pure elements were sealed in an evacuated ($1\times 10^{-6}$
torr) quartz ampoule (12mm diameter and 160mm long). The ampoule was
heated in a muffle furnace at the reaction temperature (630 $ \pm 3^0 $C
for 24 hr). The composition (Sb:Se:1:1) of the glasses was determined by
EDAX and amorphous nature was verified by X-ray diffraction pattern
obtained using by X-ray diffractometer (Philips PW 1130/60).  The
pellets of 0.677 cm diameter were prepared in the hydraulic press at a
pressure of $\approx 10^7 Pa $.
\par
        The measurements were carried out in a specially designed
vacuum cell in the temperature range 233 to 313 K and frequencies from 6
Hz to 100 kHz. A sophisticated computer controlled ac-impedance system
(Model 5206, EG \& G, PARC USA) was used for the measurements
\cite{ngprm93}. Polished samples with a thickness 0.04-0.10 cm and area
0.36 cm$^2$ were put
in contact with aquadag (a conducting emulsion) in a sandwich
configuration.  The sample was kept in vacuum in a copper sample holder
ensuring no temperature gradient between the electrods and the sample. The
effect of temperature is studied by using a closed cycle refrigerator
(Model F-70, Julabo) was used to obtain different temperatures, which can
maintain constant temperature within $\pm$ 1 K for all the measurements.
Ohmic contacts were confirmed through linear I-V characterstics in the
voltage range. All electrical measurements of real and imaginary
components of impedance parameters ($Z'$ and $Z''$) and real and imaginary
component of admittance parameters ($Y'$ and $Y''$) were made over a wide
range of temperature (233 to 313 K ) and frequency (6 Hz to 100 kHz).
\section{Results and Discussion }
\label{randd}
\subsection{Dipolar behaviour of \sbse }
\label{dpbh}
\begin{figure}
\epsfbox[0 0 200 500]{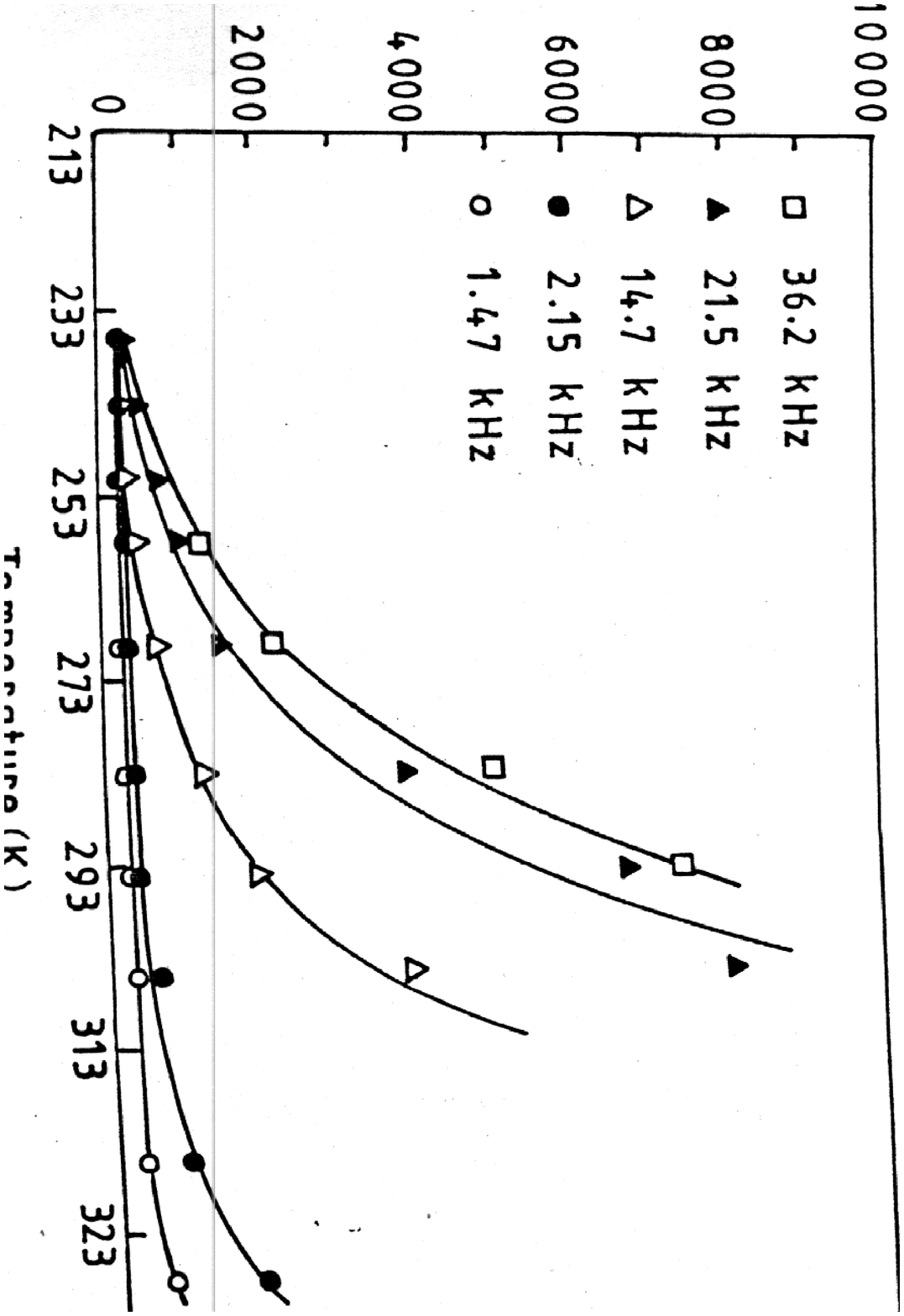}
\caption{Variation of  $\sigma_{dc} (\Omega^{-1}cm^{-1})$ with
respect to $1000/T$(/K).}
\label{f2}
\end{figure}

Figure \ref{f2} shows the plot of $\ln \sigma_{dc}$ versus 1000/T which
was obtained using the expression 
\be \sigma_{dc} = C^{\prime} \exp (-\Delta E /kT). \label{e1} \ee
The plots in Fig. \ref{f2} are found to be linear over the temperature
range studied. In the above expression $C^{\prime} = \sigma_0 \exp (
\gamma /k)$ and $\gamma$ is the temperature coefficient of the band gap.
From the figure, the activation energy $\Delta$E and $C^{\prime}$ for the
sample are found to be 0.43 eV and $ 3.35\times 10^{-4} \Omega^{-1} {\rm
cm}^{-1}$ respectively.  It has also been observed that $\Delta$E is
constant for different pellets ( thickness = 0.04cm to 0.08 cm) with a
statistical error of 0.3\%.
\par
\begin{figure}
\epsfbox[0 0 200 500]{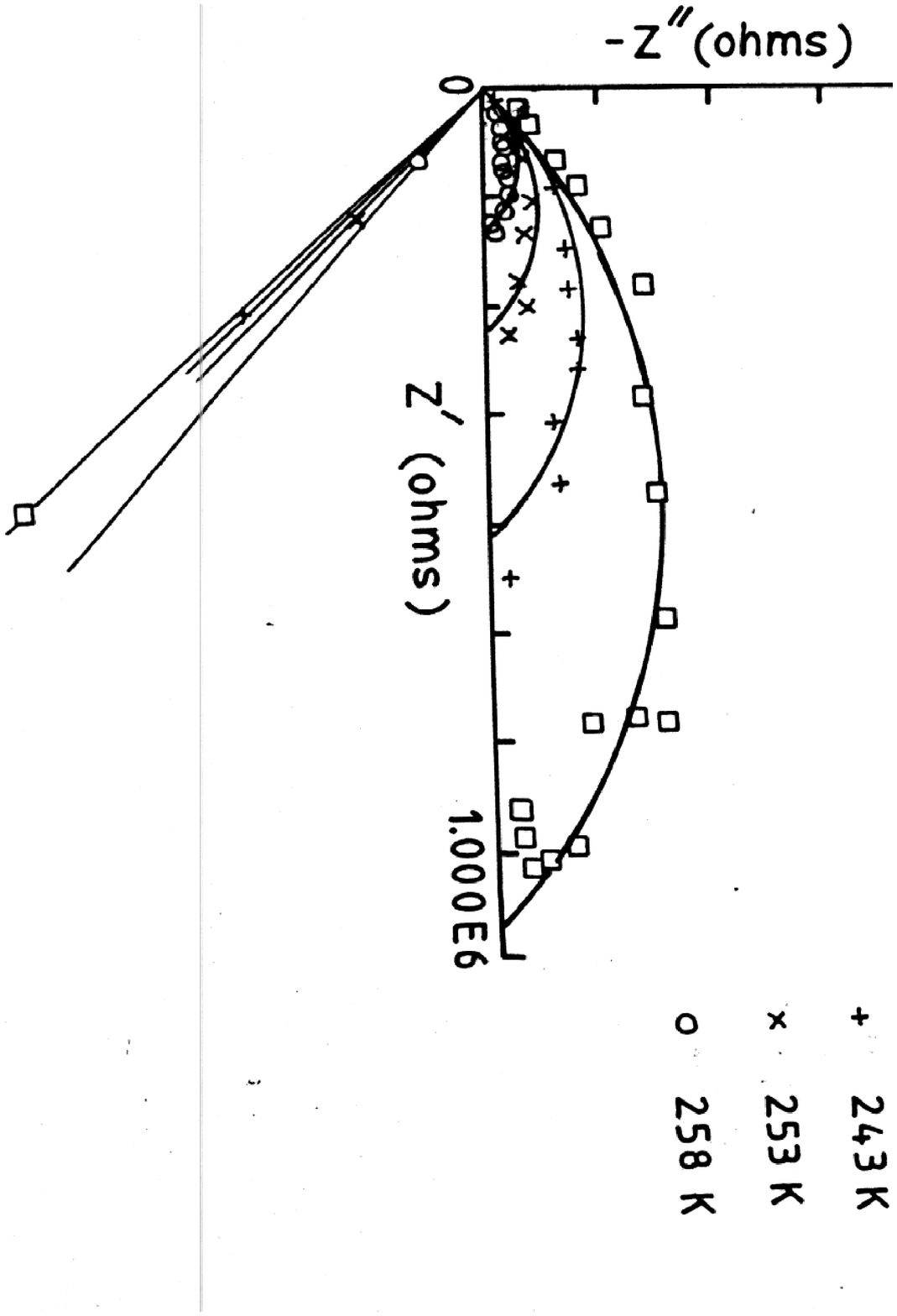}
\caption{Frequency dependence of  $\log |Z|  (\Omega)$ over the
entire temperature range studied. }
\label{f3}
\end{figure}

The variation of $\log |Z| $ versus $\log f$ (Fig. \ref{f3}) indicates
that the impedance is sensitive to frequency at lower temperatures and
gradually becomes independent of frequency with increase in temperature.
This type of behaviour indicates dominance of band conduction at higher
temperatures.
\par
\begin{figure}
\epsfbox[0 0 200 500]{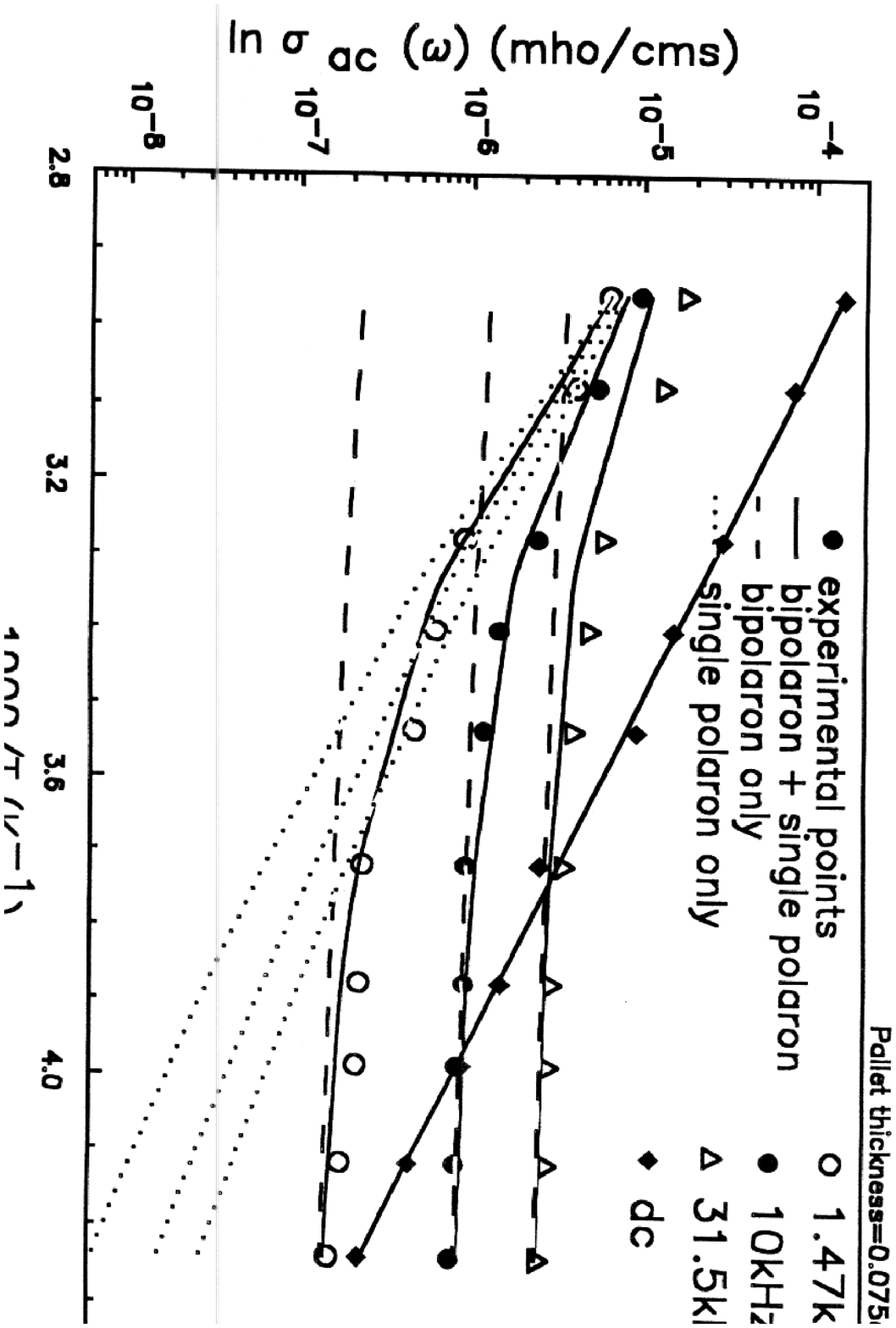}
\caption{Nyquist plots ($Z'$ versus $Z''$ ) at different
temperatures. Variation of angle $\alpha = (1-s)\pi/2$ with
respect to temperature.}
\label{f4}
\end{figure}

The sensitivity of \sbse to temperature is shown in Fig.
\ref{f4}. The figure gives the Nyquist plots ($Z'$ versus
$Z''$ ) for \sbse at different temperatures over the
frequency range studied. The semicircular Nyquist plots indicate dipolar
nature of the sample, which may be due to hopping back and forth of
bipolarons between charged defects states ( D$^+$ and D$^-$ ) reported
to be present in chalcogenides \cite{wa90}. The size of the Nyquist plots
reduces with rise in temperature  and regain their original size when
temperature is increased.
\par
\begin{figure}
\epsfbox[0 0 200 500]{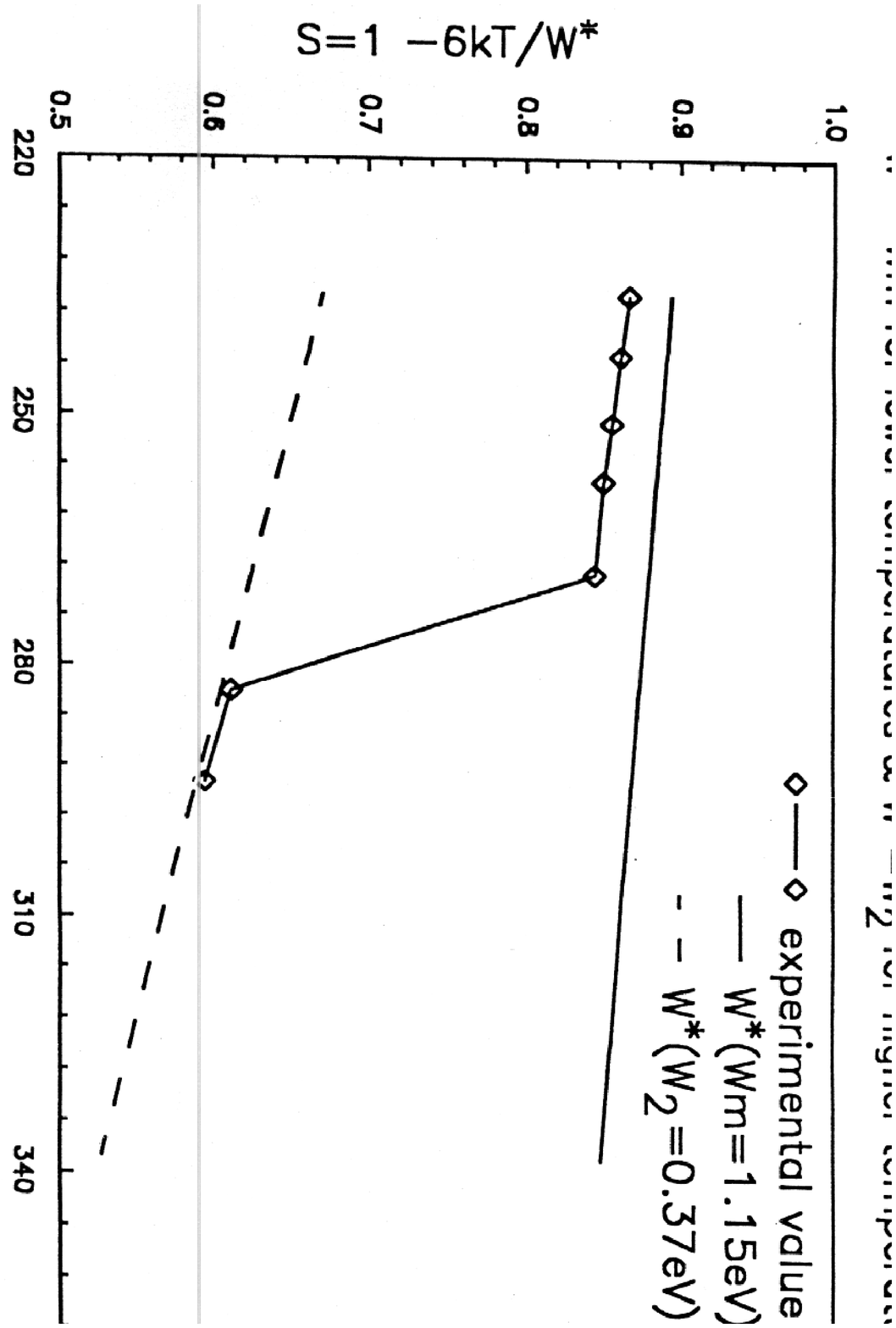}
\caption{Temperature variation of capacitance at different frequencies.}
\label{f5}
\end{figure}

The Nyquist plots for \sbse form a  perfect arc of a semicircle  with
its centre lying considerably below the abscissa (i.e., distribution
parameter $\alpha >> 0$). Figure \ref{f4} also gives the value of $\alpha$
calculated at different temperatures. In dielectric materials, the finite 
value of the distribution parameter $\alpha$ and a depressed arc are typical
for a dipolar system involving multirelaxation processes
\cite{wa90}-\cite{ngprm93}.  But the dependence of $\alpha$ on
temperature indicates that the mean position of the dipole is varying
with variation in temperature. The
dipolar nature of \sbse is further confirmed by investigating the
variation of capacitance with temperature at different frequencies. It is
clear from  Fig. \ref{f5} that in the lower temperature range the value
of C  is nearly constant and it increases with increase in temperature.
However the rate of change of capacitance (i.e., dC/dT) is higher for
lower frequencies and decreases with increasing temperature. These results
can be explained by the fact that thermally assisted hopping results in
increasing the capacitance of the material. In other words, in the lower
temperature range, the dipoles remain frozen and attain rotational
freedom when the temperature is increased. Thus the relaxation effects
are confirmed by the rate of increase in capacitance with frequency.
\par
\begin{figure}
\epsfbox[0 0 200 500]{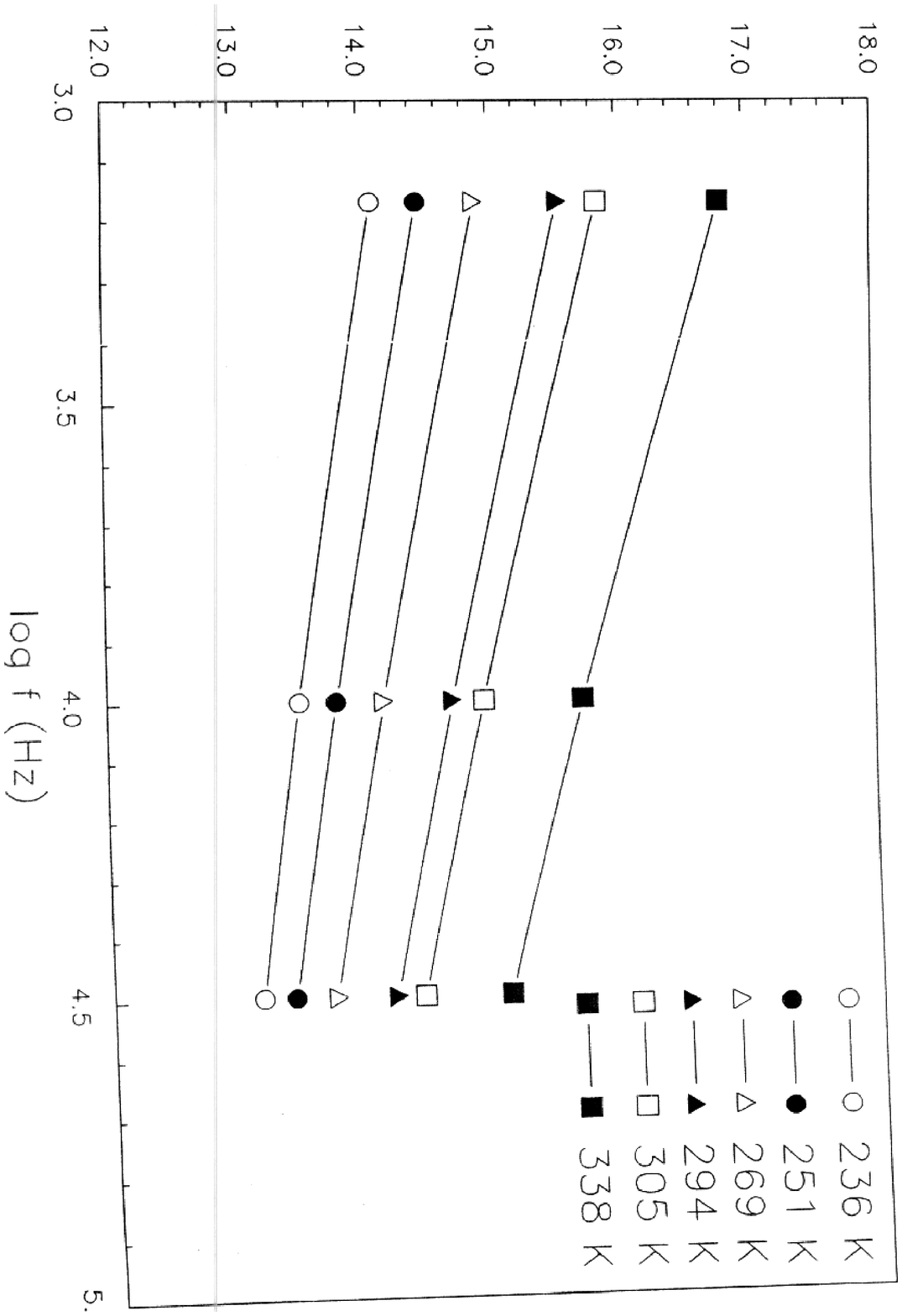}
\caption{Admittance plots ($Y' (\Omega{^-1})$ versus $Y''
(\Omega^{-1})$. The variation of angle $\alpha =s\pi/2$ with
temperature.}
\label{f6}
\end{figure}
\begin{figure}
\epsfbox[0 0 200 500]{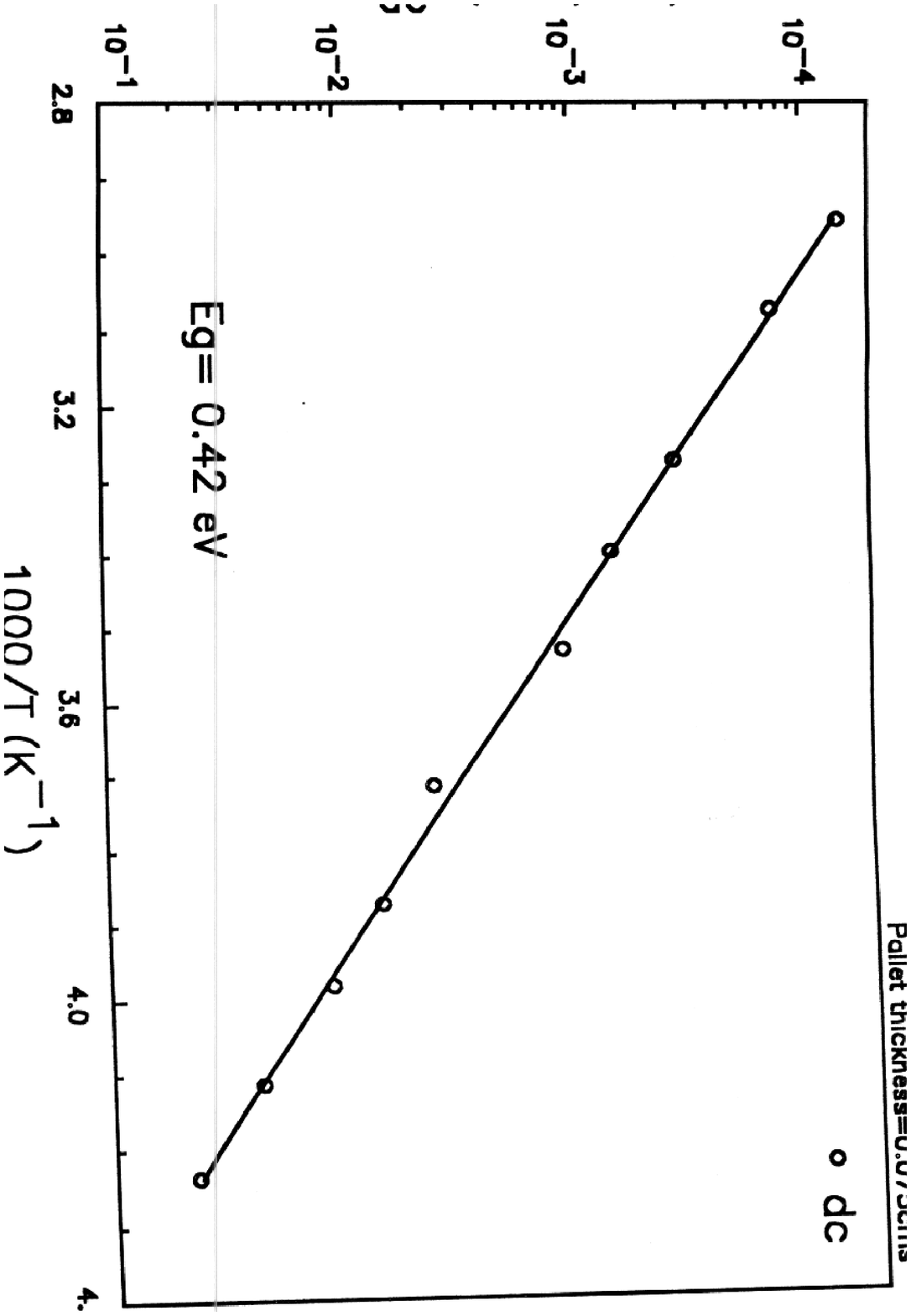}
\caption{R-C Network model for lossy capacitance.}
\label{f7}
\end{figure}

The inversion of Nyquist plots in the admittance plane are shown in the
Fig. \ref{f6}. It is clear from the figure that the angle of inclination is
less than $\pi /2$. This type of behaviour has been explained on the basis
of R-C network model (Fig. \ref{f7}), such that total impedance is given
by \cite{jonc74}
\beqn
Y (\omega) &=& B (i\omega)^n + i\omega C_{\alpha} + G_0 \num \\
           &=& G_0 + B \{ \cos(s\pi /2) + i \sin  (s\pi  /2)
           \}\omega^n + i\omega C_{\alpha}
           \label{e2} \eeqn
Moreover, this angle (shown in Fig. \ref{f6}) is also temperature
dependent. At lower temperatures (below 273 K ) it is almost constant,
while after a certain temperature (276 - 278 K) it shows a sudden
decrease, which implies  a  change  in  the conduction mechanism.
\subsection{CBH model for a-\sbse}
\label{cbh}
\begin{figure}
\epsfbox[0 0 200 500]{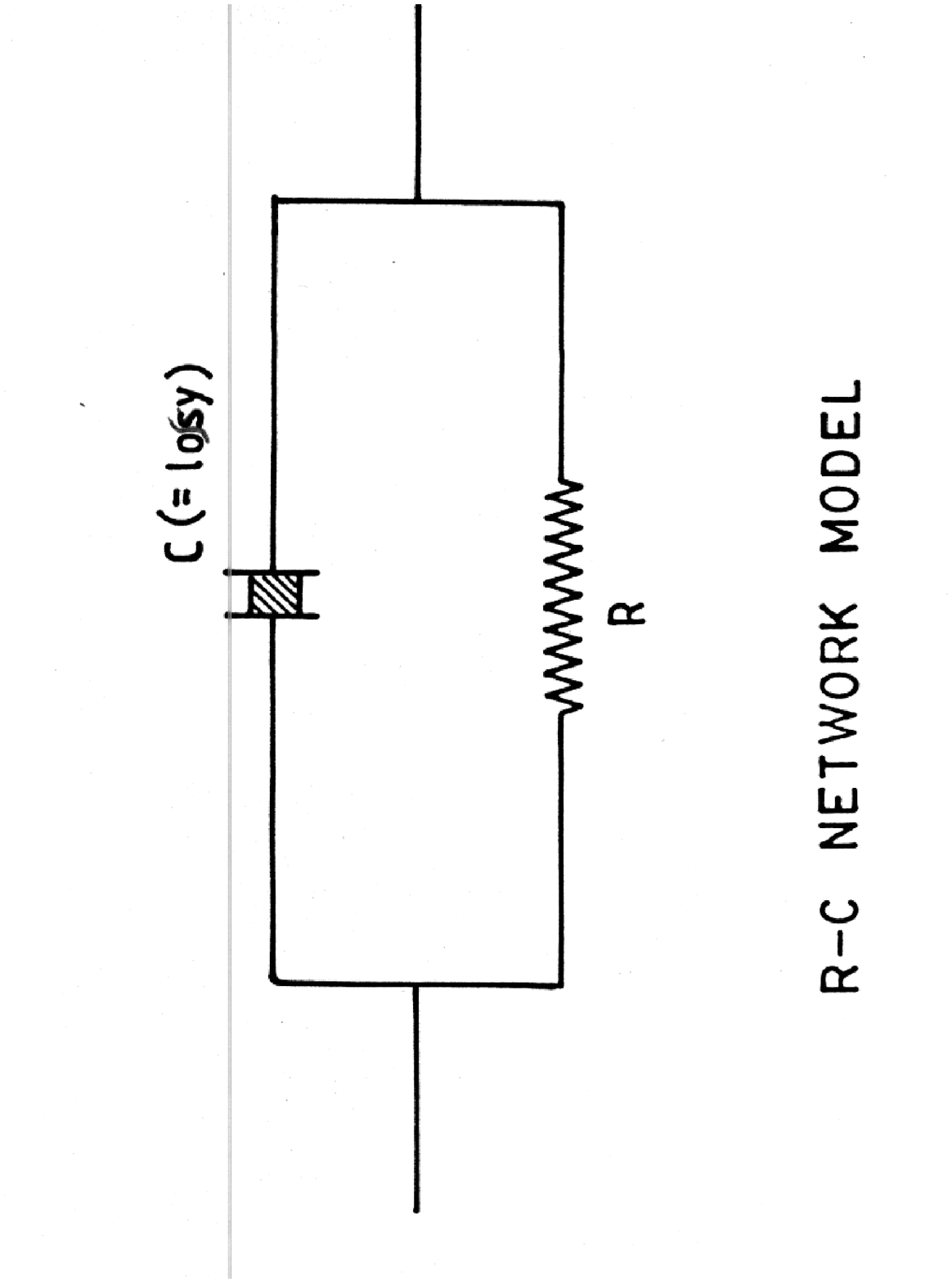}
\caption{CBH Fitting - The variation of $\sigma_{ac}
(\Omega^{-1}cm^{-1})$ with respect to $1000/T$ for experimental
points and theoretical curves (m-CBH) at different frequencies.}
\label{f8}
\end{figure}

The complex impedance plots show the dipolar multi-relaxation nature of
a-\sbse. This behaviour is explained by the R-C network model (Fig.
\ref{f7}). However, this does not further elucidate the nature of
conduction mechanism in this sample. Therefore a.c. conductivity
$\sigma_{ac} (\omega)$ at different temperatures is calculated. It is
found that dispersion in the temperature dependence of $\sigma_{ac}
(\omega)$ curves increases at lower frequencies while it merges at higher
frequencies (Fig. \ref{f8}).
\par
From  the expression
\be \sigma_{ac} (\omega) = \sigma_{total} (\omega) -\sigma_{dc}
\label{e3} \ee
the a.c. conductivity is obtained, where $\sigma_{dc}$ is given by
Eqn.\ref{e1} and its behaviour is shown in the Fig. \ref{f2}.  The
$\sigma_{total} (\omega)$ is measured directly from lock-in-analyser
(Section \ref{exp}). The $\sigma_{ac} (\omega)$ dominates over
$\sigma_{dc}$ at lower temperatures, while $\sigma_{dc}$ dominates at
higher temperatures (Fig. \ref{f8}).
\par
    The modified Correlated Barrier Hopping (m-CBH) model explains the
experimental results reported in this paper. It states that  bipolaron
hopping between the charged defects states D$^+$ and D$^-$ is responsible
for the a.c. conductivity in these semiconductors. Dipolar multi-relaxation
behaviour of this sample is already indicated from the Nyquist plots
described in the last subsection. So the charged defect states are
expected to be present in a-\sbse . According to CBH model a.c.
conductivity is given by \cite{srel77,elpm77}
\be \sigma_{ac} = \frac{n\pi^2 N N_p \epsilon' \omega
R^6_{\omega}}{24} \label{e4} \ee
where $n$ is the number of polarons involved in the hopping process,
$NN_p$ is proportional to the square of the concentration of the states
and $\epsilon'$ is the dielectric constant. $R_{\omega}$ is the
hopping distance for the conduction $\omega \tau = 1$  and is given as
\be R_{\omega} = \frac{ 4 n e^2}{\epsilon' \left[ W_m - k T \ln
(1/\omega \tau_0) \right] } \label{e5} \ee
where $W_m $ is the maximum barrier height and $\tau_0$ is the
characteristic relaxation time for the material.
\par
Figure \ref{f8} shows the variation of a.c. conductivity with
temperature at different
frequencies. It is evident from the figure that $\sigma_{ac}$ is very
sensitive to temperature in the higher temperature regime.  The low
temperature a.c. conduction can be explained by considering bipolaron hopping
between D$^+$ and D$^-$ centers whereas the higher temperature behaviour
is due to thermally  activated  single  polaron  hopping.  At  higher 
temperatures a number of thermally generated D$^0$ centers are produced
with a temperature dependent concentration \cite{elap87}
\be N_0 = N_T \exp  ( - U_{eff} /2 kT ) \label{e6} \ee
where $N_T$ is the concentration of D$^+$ or D$^-$ centers at $T = 0 K$.
The defect concentration factor $NN_P$ in Eqn. \ref{e4} is replaced by :
\beqn
N N_P = N^{2}_{T} /2 & {\rm ( For~~bipolaron~~ hopping )}\\
N N_P = N^{2}_{T} /2 ~~\exp  \left( - U_{eff}  /2kT  \right)  & {\rm
(For~~  single~~  polaron~~ hopping ) }  \label{e7} \eeqn
Thus for single bipolaron hopping this factor is thermally activated and
hence $\sigma_{ac} (\omega)$ is also activated.
\par
\begin{figure}
\epsfbox[0 0 200 500]{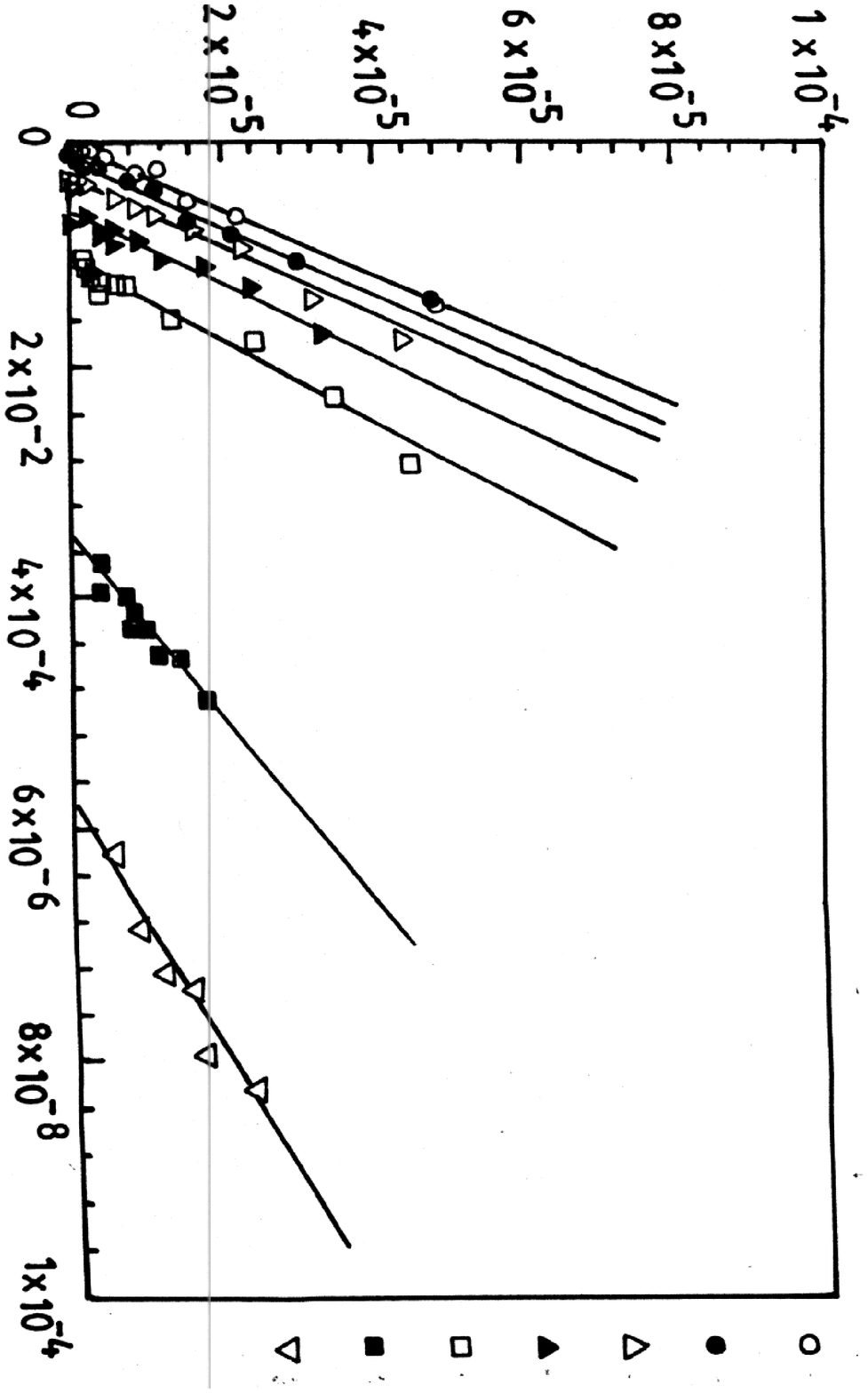}
\caption{Temperature dependence of $s (=1-6kT/W^*)$ for
theoretical curves (m-CBH) and experimental points.}
\label{f9}
\end{figure}

According to CBH model \cite{shim82}
\be B =W_m -W_1 +W_2 \label{e8} \ee
where $B$ is optical band gap, $W_m$ is the maximum band width and
behaviour with frequency $\omega$ is as 
\be \sigma_{ac} (\omega) = A \omega^s \label{e9} \ee
Here [\cite{srel77}],
\be s =\frac{d(\ln \sigma_{ac})}{d \ln \omega} =1 - 6kT/W   \label{e10}
\ee
While $W=W_m$ for bipolaron hopping and $ W = W_1~ or~ W_2$ for single
polaron hopping. Therefore, $W_m$ is estimated from the $s$ values at
lower temperatures. Values of $W_1$, $W_2$ and $U_{eff}$ were estimated
from the values of $s$ at higher temperatures. These parameters are then
fed to CBH model to fit the a.c. conductivity data calculated from the
experiments. The behaviour of Eqn.\ref{e10} ($s = 1 - 6kT/W^*$) can be
seen in the Fig. \ref{f9}. This indicates that $W^*$ follows $W_M$ for
lower temperatures and $W_2$ for higher temperatures.
Figure \ref{f9} also shows a sudden change in conduction mechanism at a
certain temperature (276-278K).
The similar behaviour is noticed in the admittance plots (Fig. \ref{f6}).
The different parameters used for CBH model calculations are as following

\begin{figure}
\epsfbox[0 0 200 500]{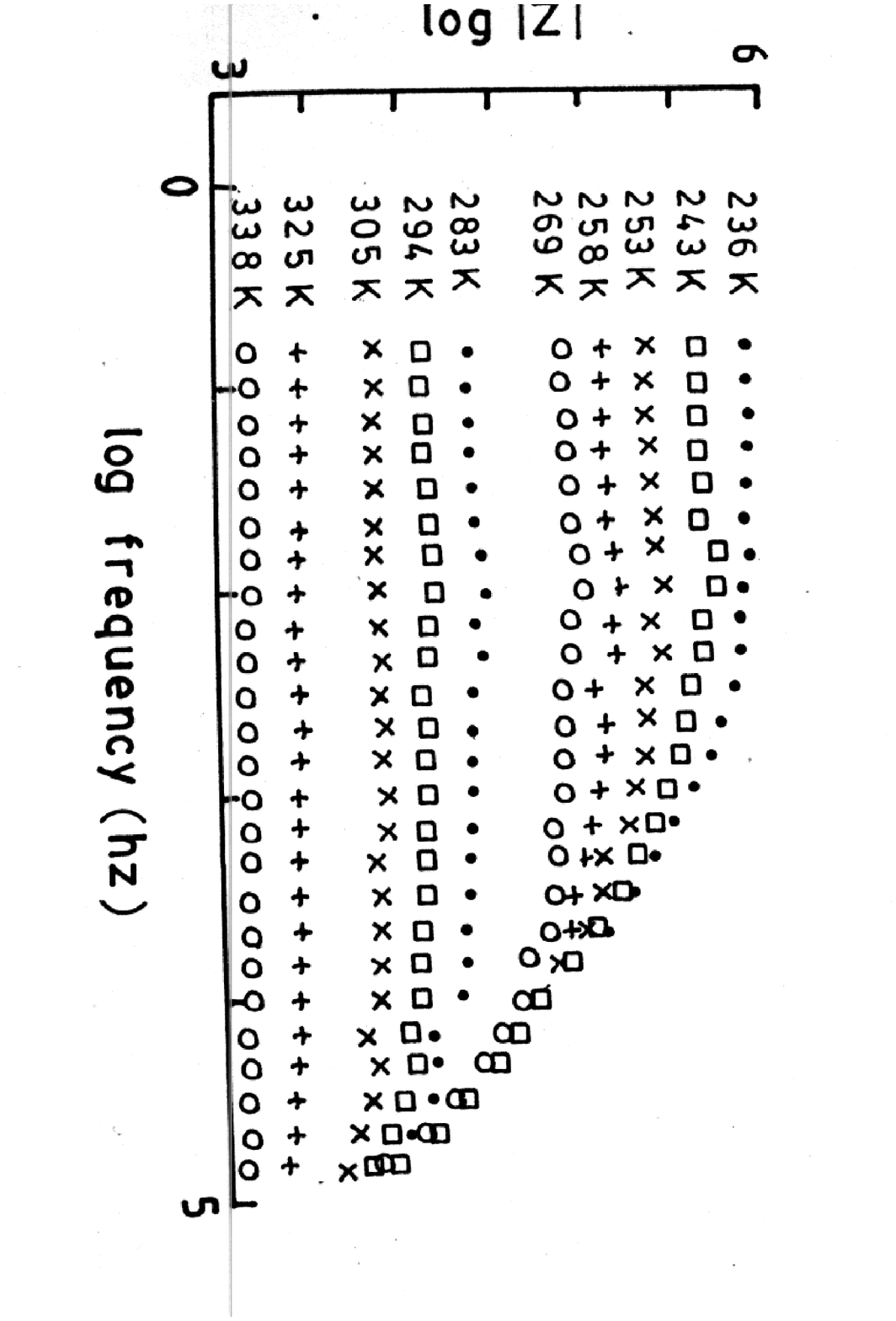}
\caption{Variation of $R_{\omega} (A^0)$ with respect to
temperature (K) at different
frequencies}
\label{f10}
\end{figure}

Further, the effect of temperature and frequency on the hopping length
$R_{\omega} $, which is a measure of the effective length of a dipole can be
studied using CBH model. The variation of $R_{\omega}$ with frequency at
different temperature is shown in the Fig. \ref{f10}. The figure indicates
that $R_{\omega } $ is more sensitive to frequency in the high temperature
regime and the sensitivity decreases with decreasing temperature.  Thus at
low temperature, $R_{\omega}$ is constant with frequency, which implies a
constant value of capacitance which is found to be true from the Fig.
\ref{f5}.

\section{Conclusion }
\label{con}
Nyquist plots or complex impedance studies confirm the dipolar nature of
\sbse and multi-relaxation behaviour as seen in most chalcogenide
glasses.  These plots also indicate that the capacitive nature dominates over
the resistive nature of the sample at lower temperature ( $< 273$ K ),
while the resistive nature is dominant at higher temperatures. The conduction
mechanism has a sudden change from bipolaron to single polaron hopping
(276-278 K). The temperature and frequency dependence of a.c. conductivity
is well explained by the m-CBH model. The contribution to a.c.
conductivity from single polaron hopping is dominant at higher temperatures.
\subsection*{Acknowledgements}
The authors are grateful to University Grant Commission for providing
funds under COSIST and CAS programmes for purchase of measurement system
which was essential for investigations reported in this paper. 

\listoffigures
\vfill\eject

\end{document}